# The E-ELT Multi-Object Spectrograph: latest news from MOSAIC

F. Hammer[a], S. Morris[b], L. Kaper[c], B. Barbuy[d], J. G. Cuby[e], M. Roth[f], P. Jagourel[a], C. J. Evans[g], M. Puech[a], E. Fitzsimons[g], G. Dalton[h,i], M. Rodrigues[a], and the MOSAIC team

[a]GEPI, Observatoire de Paris, CNRS, Univ. Paris Diderot, Place Jules Janssen, 92190 Meudon, France; [b]Department of Physics, Durham University, South Road, Durham, DH1 3LE, UK; [c]Astronomical Institute Anton Pannekoek, Amsterdam University, Science Park 904, 1098 XH, Amsterdam, The Netherlands; [d]Universidade de Sao Paulo, IAG, Rua do Mato 1226, Cidade Universitria, Sao Paulo, 05508-900; [e]Aix Marseille University, CNRS, LAM UMR 7326, 13388, Marseille, France; [f]Leibniz-Institut für Astrophysik Potsdam (AIP), An der Sternwarte 16, 14482 Potsdam, Germany; [g]UK Astronomy Technology Centre, Royal Observatory Edinburgh, Blackford Hill, Edinburgh, EH9 3HJ, UK; [h]Astrophysics, Department of Physics, Keble Road, Oxford OX1 3RH; [i]Space Science and Technology, Rutherford Appleton Laboratory, HSIC, Didcot OX11 0QXS

**ABSTRACT**

There are 8000 galaxies, including 1600 at z≥ 1.6, which could be simultaneously observed in an E-ELT field of view of 40 arcmin$^2$. A considerable fraction of astrophysical discoveries require large statistical samples, which can only be obtained with multi-object spectrographs (MOS). MOSAIC will provide a vast discovery space, enabled by a multiplex of 200 and spectral resolving powers of R=5000 and 20000. MOSAIC will also offer the unique capability of more than 10 `high-definition' (multi-object adaptive optics, MOAO) integral-field units, optimised to investigate the physics of the sources of reionization. The combination of these modes will make MOSAIC the world-leading MOS facility, contributing to all fields of contemporary astronomy, from extra-solar planets, to the study of the halo of the Milky Way and its satellites, and from resolved stellar populations in nearby galaxies out to observations of the earliest 'first-light' structures in the Universe. It will also study the distribution of the dark and ordinary matter at all scales and epochs of the Universe.
Recent studies of critical technical issues such as sky-background subtraction and MOAO have demonstrated that such a MOS is feasible with state-of-the-art technology and techniques. Current studies of the MOSAIC team include further trade-offs on the wavelength coverage, a solution for compensating for the non-telecentric new design of the telescope, and tests of the saturation of skylines especially in the near-IR bands. In the 2020s the E-ELT will become the world's largest optical/IR telescope, and we argue that it has to be equipped as soon as possible with a MOS to provide the most efficient, and likely the best way to follow-up on James Webb Space Telescope (JWST) observations.

## 1. INTRODUCTION

The European Southern Observatory (ESO) has signed a phase A study contract with the MOSAIC consortium that includes five countries (France, UK, The Netherlands, Brazil and Germany); another six countries are Associated Partners (Austria, Finland, Italy, Portugal, Spain, and Sweden). The phase A study goal is to develop two instrument concepts, one limited by the hardware budget provided by ESO (18 M€), an alternative one generating a sufficiently efficient and competitive MOS as delineated by the top level requirements (TLRs) defined by [1] and by the E-ELT Project Science Team. The MOS capabilities will be unique in the 2020s (see Figure 1) making an efficient trade-off very challenging, especially when one accounts for (1) the necessary simplicity of the MOS concept and (2) the difficulty to predict what the science priorities will be during the 2020s.

We have initiated a trade-off study to account for:
- <u>Field of view:</u> the redesign of the secondary mirror (when the telescope size was reduced from 42 to 39m) implies that incident light is now not perpendicular to the focal plane (non-telecentric), which places strong constraints on the architecture for selecting science targets;

- <u>Instrument complexity and cost:</u> keeping the facility able to address many TLRs (visible and NIR coverage, high-definition and high-multiplex modes, GLAO-fed IFUs for IGM tomography) requires a unique system of fibres to cover the focal plane;

- <u>Blue transmission:</u> For the best overall performance over the visible/IR domain, the E-ELT mirrors will be coated with a combination of silver and aluminium (Ag+Al). This gives a significant gain in throughput from the (five-mirror) telescope compared to bare aluminium, but suffers from poor reflectivity shortwards of ~0.4 μm;

- <u>Competition with other facilities:</u> the 2018 launch of the James-Web Space Telescope (JWST) requires that we make performance comparisons, especially relatively to the background at the reddest wavelengths.

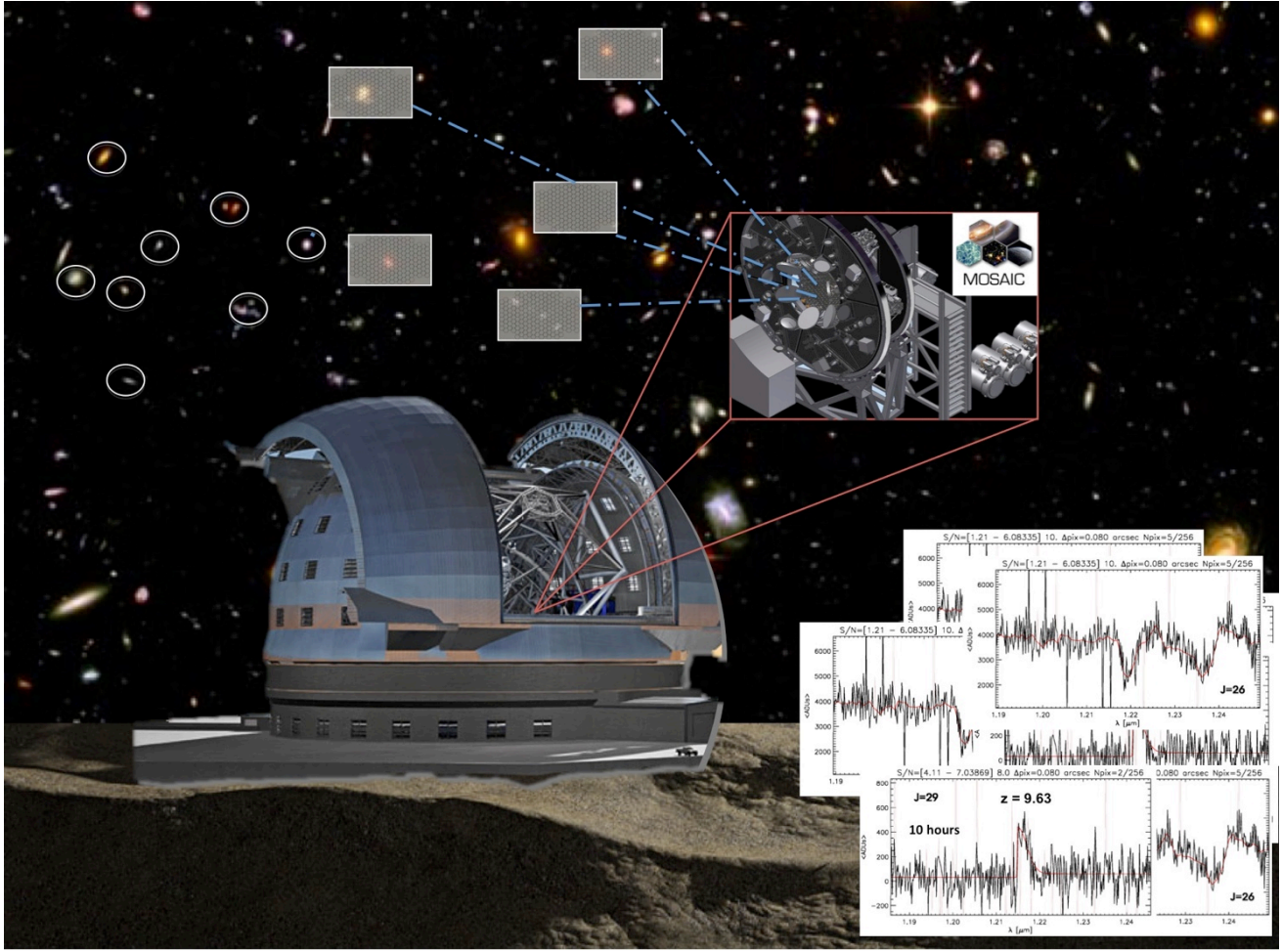

Figure 1. A montage to illustrate how MOSAIC, when implemented on the E-ELT, will take hundreds of spectra exploiting its HDM-IFUs (rectangles) and HMM-fibres (circles), allowing the discovery and the quantitative study of the faintest and most-distant galaxies in the Universe.

Table 1 lists the preliminary TLRs of the HDM and HMM modes (but not the 'IGM' mode that is made of visible IFUs), in which, e.g., the multiplex has already decreased compared to the ESO TLRs based on the limited hardware budget. Although quite minimal for an E-ELT MOS (e.g., there are 8000 $H_{AB} \leq 28$ galaxies in a 40 arcmin E-ELT field of view), it is very ambitious given the numerous constraints described above. Fortunately MOSAIC is supported by a very active (currently 127 members) and enthusiastic Science Team, and the E-ELT MOS science cases cover a large fraction of the future scientific prospects that have motivated the construction of the E-ELT, with the promise to revolutionize our current understanding of the Universe.

We are in the process of achieving a full architecture of MOSAIC as well as its operational modes. During mid-Phase A this will lead to an estimate of what the MOS budget should be to make sure that the E-ELT will be at the forefront in cosmology, extragalactic and stellar astronomy.

Table 1. The initial trade-offs for the MOSAIC Phase A study (HDM: High Definition Mode; HMM: High Multiplex Mode).

| Parameter | Value | Tolerance | Comments |
|---|---|---|---|
| **Aperture: HDM** <br><br> **HMM** | Square/Rect 2.0"x 2.0" <br> Two apertures: <br> 0.6" in NIR and <br> 0.9" in visible | The values are minimum requirements. | No trade-off in the size or shape possible. <br> Assuming GLAO performance in NIR (0.8" if seeing-limited). *Trade-off GLAO/ seeing limited in NIR*. |
| **Multiplex: HDM** <br> **HMM** | 10 <br> 200 | Minimum: 10 <br> Minimum: 200 | *Could be lower if using cross-beam switching to optimise sky subtraction* |
| **Spatial pixel size (HDM)** | 75 mas | Minimum: 40 mas <br> Maximum: 80 mas | |
| **Ensquared energy (HDM)** | 30% | Minimum: 25% | In 2x2 spaxels in H band <br> in the full MOSAIC field-of-view |
| **Spectral resolution:** <br> **HDM** <br> **HMM** | 5000 <br> Two modes: 5000 and 15000 (mandatory in Vis, desirable in NIR) | Minimum requirement values for both NIR and visible | |
| **Total coverage: HDM** <br><br><br> **HMM** | 0.8 to 2.5 μm <br><br><br> 0.4 to 1.8 μm | Mandatory 1.0 to 1.8μm, desirable 2.5 μm <br> Desirable 0.38 to 2.5 μm | The wavelength coverage in which MOSAIC will operate under MOAO conditions. *Trade-off on the upper limit for HDM: 1.8 or 2.5 μm.* |
| **Simultaneous coverage: both HDM & HMM** | One broad band at R=5000 | · | The simultaneous coverage is not needed to be equal in each band. Bluer settings will have smaller coverage. |
| **Science field-of-view both HDM & HMM** | 40 arcmin² | Minimum: 40 arcmin² <br> Maximum: 78 arcmin² | Maximize patrol field |

## 2. LATEST SCIENCE NEWS

### 2.1 Update of the MOSAIC Science Cases

The number and significance of the MOSAIC Science Cases has increased much during the last few years since the conclusions of the, EAGLE, OPTIMOS-EVE and OPTIMOS-DIORAMAS phase A studies; here we summarize the most important developments:

· *'First light':* MOSAIC will detect and study the very first galaxies. Their light, which has taken more than 13 billion years to reach us, will provide us with vital clues to our understanding of the early epoch when the Universe was 'reionised', during which its gas changed from a universally neutral into an ionised state.

- *'Mapping the inter-galactic medium':* The warm and hot gas between galaxies and within their halo is a reservoir of matter from which proto-galaxies can form, or which can 'feed' gas into existing galaxies, providing the raw material for new generations of stars. MOSAIC will provide an unprecedented map of the distant 3D structures of this gas.
- *'Galaxy evolution with cosmic time':* MOSAIC will dissect galaxies over the full lifetime of the observable Universe, providing us with vital information on their physical and chemical properties. The MOSAIC HDM will also provide well sampled rotation curves measuring the dark matter content of massive galaxies up to z=4. MOSAIC will be unrivalled when studying lower mass 'dwarf' and low-surface brightness galaxies, which play a major role in shaping galaxy evolution in the frame of the hierarchical scenario.
- *'Super-massive black holes':* A key question for astronomers is how the growth of super-massive black holes (thought to be at the heart of most present-day galaxies) and their host galaxies are self-regulated by feedback processes. These could be related to massive outflows driven by active galactic nuclei (AGN) and supernova explosions. MOSAIC will provide the first meaningful samples of galaxies in which the physical and geometrical parameters of such outflows will be measured.
- *'Stellar populations in the Milky Way and beyond':* The evolutionary history of galaxies is imprinted on their stellar populations, via their ages, chemical abundances, and kinematics. Only with spectroscopy we obtain robust estimates of these properties, to confront theoretical models of galaxy evolution. MOSAIC will study the oldest stellar populations in the Milky Way and nearby galaxies, providing a unique and direct connection to the physical conditions in the first galaxies. It will also observe individual stars in galaxies out to tens of Mpc, exploring an unprecedented range in stellar environments and providing direct estimates of chemical abundances for a large volume of the local Universe.
- *'Exploring the centre of the Milky Way':* One of the most spectacular results of the past decade was arguably the observed orbits of stars aroud Sgr A*, the massive black hole at the centre of our Galaxy. Surrounding this central region there are puzzling structures of gas, dust, and sites of associated star formation, but these remain out of reach of current facilities. MOSAIC will provide the first glimpse into the physical conditions in this elusive region.
- *'Planet formation in different environments':* The number of exo-planets known is growing rapidly, but this is posing important questions regarding the importance of environment – specifically stellar density and metallicity – on their formation. MOSAIC will be able to undertake comprehensive radial-velocity studies of stars in considerably more diverse environments than currently possible, e.g., in open and globular clusters, spanning a wide spread of densities and metallicities, at a range of distances from the centre of the Galaxy.

The first three science cases will have a strong impact on cosmology, including the formation of the first structures, the missing baryon problem and the evolution of dark matter from well established rotation curves up to z= 4.

Activities on the science side have mostly been devoted to science trade-offs and setting up the simulation tools that will be used to develop the MOSAIC Reference Surveys.

**2.2 Trade-off regarding extension of MOSAIC to include K-band**

A long-standing discussion in defining the top-level requirements of MOSAIC is the potential inclusion of the K band (1.95-2.45 μm). In [1] we have summarized the scientific motivation argued by the MOSAIC Science team to include this band in the proposed 'high-definition' mode making use of MOAO.

The primary motivation is to observe the rest-frame UV and optical features of very high-redshift galaxies, but there are also demands on the use of the K band in Galactic Science cases. The requirement for K-band observations with the high-definition mode of MOSAIC is mostly driven by the extragalactic cases, and most strongly by the galaxy assembly/dynamics case for which extensive simulations were conducted during the E-ELT DRM [2]. These simulations already revealed the dramatic impact of the telescope thermal background in the K band (see Figure 2) that hamper decent signal-to-noise ratios to be obtained on the most distant galaxies in reasonable exposure times. However, not including the K band into the MOSAIC design would reduce our ability to study the chemo-dynamics of distant galaxies, including dynamics and metallicity gradients of z > 4 galaxies, and not the least, the possibility to detect HeII/CIII] at z ~ 10-12.

To evaluate the impact of the K band, we need to compare the E-ELT performance with that of JWST in particular, on which NIRSPEC will be equipped with multi-slits as well as with a single IFU. Figure 2 shows that above 2 μm, the 39m telescope thermal background increasingly dominates, which rapidly reaches 1000 times the zodiacal light that limits JWST's performance in space. We are carrying out an analytic comparison of the MOSAIC HDM vs. the JWST/NIRSpec single IFU (see more details in [3]). These preliminary calculations reveal that for J- and H-band observations MOSAIC with 10 IFUs in the HDM mode will be a few times more efficient than assembling surveys of targets one at a time with NIRSpec. Observations at 2.15 μm will not be competitive at the low spectral resolution of NIRSpec (bearing in mind the uncertainties associated with such calculations). The significant thermal background of the

E-ELT at longer wavelengths (2.4 μm) means that MOSAIC loses any advantage with respect to JWST (assuming that NIRSpec will be limited by zodiacal light).

Refinement of the analytical calculations are on-going, taking into account the impact of the detectors (dark current for JWST, readout noise for the E-ELT), as well as inclusion of a comparison with HARMONI[1].

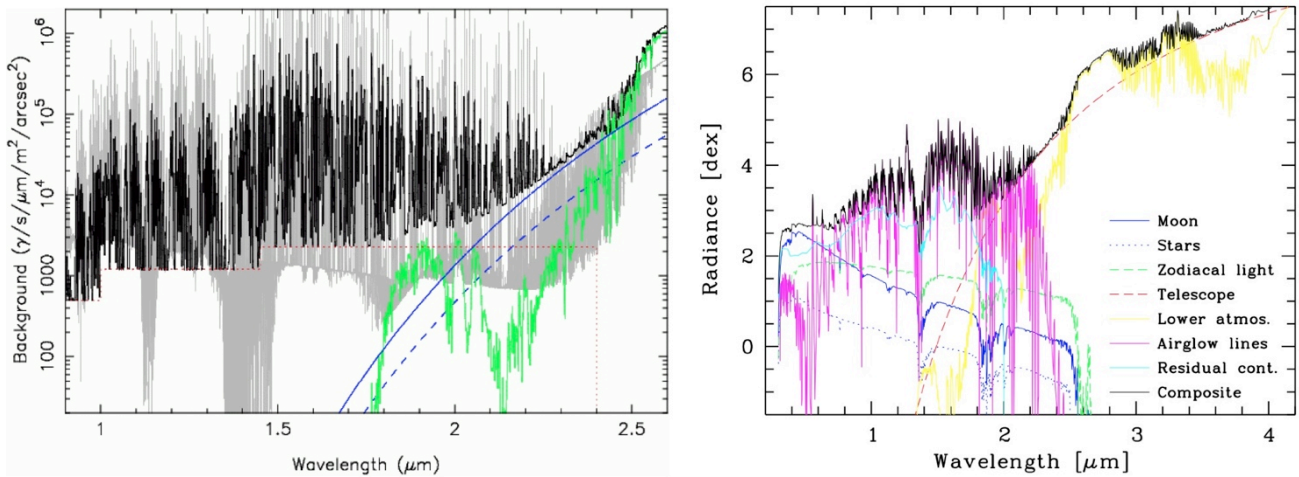

**Figure 3:** *(Left)* Comparison between different sources of background flux for the E-ELT[2]. *Black:* Near-IR background model for a Paranal-like site (airmass = 1, bare Al coating, *R*=1000); *Red (dotted):* representative continuum, assumed to be constant in each band, and which does not extend beyond the *K*-band. *Blue:* thermal background from the telescope assuming bare Al (solid line) and Al+Ag (dashed line) coatings. *Green:* thermal emission from the atmosphere. *Grey:* near-IR background model (at greater spectral resolution) from the Gemini Exposure Time Calculator for Mauna Kea for comparison. *(Right)* Components of the Paranal sky model in logarithmic radiance units for wavelengths between 0.3 and 4.2 μm, during a night with the moon over the horizon (from [4]). The scattered light and airglow components were only computed up to the K band because of their negligible importance at longer wavelengths.

## 2.3 Analysis of potential blue-visible performance of MOSAIC

The potential performance of MOSAIC at the blue end of the visible spectrum will be investigated (see also [3]). This is primarily motivated by a couple of the science cases (IGM tomography and Galaxy archaeology).

The primary consideration/concern regarding the blue-visible performance of E-ELT + MOSAIC is the throughput of the telescope itself due to the coating of the E-ELT mirrors (silver and aluminium, Ag+Al) that suffers from poor reflectivity (<20%) short-wards of ~0.4 μm. Based on the preliminary design and performance estimates, the total throughput of E-ELT+MOSAIC at 0.4 μm is estimated to range between 4-7 % depending on how the spectrograph is optimised (i.e., blue-optimised vs. red-optimised design). The study will also include a relative comparison of the expected signal-to-noise ratio of MOSAIC vs. current instrumentation at the VLT working at ~0.4 μm, and other ELTs.

Our preliminary conclusion is that MOSAIC observations at 0.37-0.40 μm are outperforming with current VLT instrumentation, but that any new blue-optimised MOS on the VLT could be competitive. At this stage in Phase A we recommend that we continue to investigate the performance of potential designs in the blue, keeping a careful eye on the estimated end-to-end throughputs to ensure that they remain competitive. In particular, if the spectrograph is (necessarily) optimised for the red-visible then the case for UV coverage ≤ 0.40 μm would be significantly weakened.

---

[1] Our results are preliminary since they do not seem to be consistent with results showing that the single HARMONI IFU might be more efficient than

[2] From the ESO E-ELT DRM technical data repository: https://www.eso.org/sci/facilities/eelt/science/drm/tech_data/background/

### 2.4 Simulation Work Package

Activities in this WP focused on the set up of the simulation tools to be used to further test Science Cases and to develop MOSAIC Reference Surveys. An end-to-end simulator [5] that includes expected performances of the telescope and/or the MOSAIC AO system has been developed, building on the COMPASS project effort led by LESIA and GEPI over the past two years. Compared to the tools used during the previous phase A studies, the new WEBSIM-COMPASS simulator (http://websim-compass.obspm.fr/) includes refined capabilities (updated PSFs, inclusion of LSF and differential atmospheric refraction). It will also allow the MOSAIC science team to run simulations in *batch* mode, i.e., to run a large number of simulations at the same time. This will be required to explore a large parameter spaces during the development of the MOSAIC Reference Surveys. The release of the WEBSIM-COMPASS simulator to the Science Team is expected mid-July 2016.

### 2.5 Interaction with the Science Team and Project Science Team (PST)

Beyond the trade-off analysis and simulation activity, a MOSAIC Science meeting was held in Paris in March 2016 just before the MOSAIC Kick-Off Meeting. The instrument Top Level Requirements were reviewed and feedback from the MOSAIC Science Team was collected. Responses to actions from the E-ELT PST on the requirements for the ELT-MOS were collected and sent to ESO.

We acknowledge the PST for their constructive input and thoughts at this early stage of the design process and we are eager to maintain this interaction during the course of the Phase A study.

## 3. MOSAIC SYSTEM

This section summarises the current status of the technical work, which is currently focused on architecture trade-off.

### 3.1 K-band feasibility

Key to the MOSAIC multi-mode operation is the shared spectrograph concept, in which the spectrograph (optics and detector) serves both the 'HMM' and 'HDM' or 'IGM' modes. In particular, this presents an issue for K-band operation, which is listed as a goal for the HDM mode in the current TLRs.

The HMM mode is only feasible with a fibre-fed spectrograph due to the high multiplex (200), therefore operation at K-band with IFUs utilising the same spectrographs will be significantly more complex/expensive unless fibres are also used for the HDM mode. An assessment on using fibre-optics in the K-band was performed, based on existing experience and available technology. The conclusion is that the prospects of operating a fibre-fed spectrograph in the K band are poor, with available fibres which operate in the K band being expensive and typically only available in short lengths or suffering from excessive modal noise.

In parallel, the capabilities of MOSAIC in the K band were re-assessed by the science team as described above. A TLR validation exercise will be performed, consolidating the work that has been performed so far including the K-band analysis. The results of this exercise are pending, but it is envisioned that the K band might have to be dropped from the baseline, mostly for reasons of cost.

### 3.2 Adaptive Optics

A dedicated kick-off meeting for the adaptive optics trade-off analysis was held at the beginning of April 2016. As part of this, a comprehensive set of baseline parameters for AO simulations has been formulated to ensure consistency and interoperability between the various simulator environments being used (at Durham, LESIA, ONERA and LAM). These parameters have been documented and cover:
- A set of baseline parameters and simulations to allow direct comparison between the various simulators;
- A range of parameters derived from the science requirements and ESO specifications;
- A number of parameters where the range is defined by hardware availability and costs (such as µDMs).

In addition, four main architecture trade-offs were identified which will be prioritised due to their impact on the rest of the system, namely: the number of LGS to be used (0, 4, and 6 will be investigated), whether or not to use MEMS DMs, whether the LGS should track the pupil or the sky, and an assessment of the AO performance with GLAO only.

Discussion is also on-going to identify the most representative fields for testing AO configurations in simulation. So far, three clarification requests to ESO have been submitted.

### 3.3 Pupil Sheer

The allowable offset of the pupil on a fibre end – the pupil sheer – is a critical parameter. It is not feasible to inject a pupil that is exactly the same size as the fibre core, as various manufacturing and alignment tolerances, plus telescope stability, will lead to offsets of the pupil from the centre. This can cause variable clipping that will render any flat-field calibration inaccurate. Instead, the pupil must be over- or undersized with respect to the fibre core such that the throughput remains constant.

A budget for the allowable offset of the pupil, taking into account all known sources of pupil offset at present, has been developed. With a substantial technical immaturity margin applied, an over (or under)sizing of the pupil by 20% will be targeted. Whether an over or undersize will be used is still to be determined, and will depend on the required throughput (which is less for an oversized pupil), the size of the spectrograph optics (which increase for an undersized pupil), and the manufacturability of the fibre (which decreases for an oversized pupil). We are in the process to model these and other possible effects (see, e.g., [6], and references therein).

### 3.4 NIR architecture

Work has also started on the optical architecture for the NIR-channel (HDM and HMM modes). This is considered to be more difficult than the visible channel due to the large mismatch in spaxel size between the two modes to be shared by the spectrograph (ca. 75 mas and 200 mas, respectively).

A model of the optical chain is being developed which allows for a quick investigation of changes in various design parameters. A key difficulty is, with no other action taken, having a slit of 75 mas and 200 mas sharing the same spectrograph will automatically mean that, for a spectrograph sized for the narrower slit, the wider-slit mode (HMM) will unavoidably be a) oversampled on the detector and b) have higher spectral resolution when using the same grating. Since the ideal is for both modes to have the same (or similar) spectral resolution and sampling on the detector, the only solution is to change the size of the slit, if a truly shared mode spectrograph is to be employed.

A concept has been developed which involves slicing the HMM ~ 200 mas slit by three. In initial calculations, this allows a single spectrograph (i.e. collimator, grating, camera and detector) to image both the HDM and HMM slit with the desired properties and avoiding the need for overly large optics or an extremely fast camera. Two options for how to achieve the slit slicing have been identified, and both are being investigated further. They are: firstly utilising an image slicer at the slit output: this could either be a slicer per fibre, per object, per section of the slit or per slit; and secondly using a spliced fibre in which a larger input fibre is spliced into three smaller fibres (cf. how a fibre laser works). An initial discussion with a manufacturer indicates that this is likely feasible with the rough fibre sizes being considered but further investigation is required. With this concept, no slit-exchange mechanism would be used – instead the slit images would be placed close to each other and an actuated mask would be used to cover the slit not being used. This constitutes a useful simplification, as the requirements on the slit mask exchange would be much simpler than on a slit-exchange. The penalty would be the need for slightly larger optics and a $\Delta\lambda$ offset on the detector (that could be easily calibrated out).

Finally, some thought has been given to the requirement to have a high spectral resolution mode (R=15000?) in the visible – and as an option in the NIR. Since MOSAIC will almost certainly only image a section of the total spectral range in one exposure, a precision rotating grating is likely required. This will greatly complicate the adoption of a high-resolution mode as a simple grating exchange mechanism (e.g. a wheel) cannot be employed, and likely a re-configurable optical path system will be required. This will be investigated further, but it is suspected that this issue will complicate the incorporation of a high-resolution mode.

## 3.5 Focal plane, positioner and the non-telecentric telescope

The non-telecentricity of the telescope is greatly complicating the design of the focal plate. The 'stepped and tiled' solution identified before the start of the Phase A study (see Figure 3) remains the favoured solution and a trade-off of possible actuation technologies is on-going.

Initial calculations have indicated that for a 30 arcsecond diameter tile (i.e. a tile with a 1 arcminute patrol field), defocus even at the edge of the field would degrade the PSF by less than 0.1 arcsecond. This is only relevant for the HMM mode (HDM and IGM modes require separate focus compensation anyway), but it is highly likely that a 0.1 arcsec blurring can be accommodated passively, removing the need for additional focus compensation on each tile. Confirmation of this, however, is pending on the AO analysis on seeing-limited and GLAO-based operation.

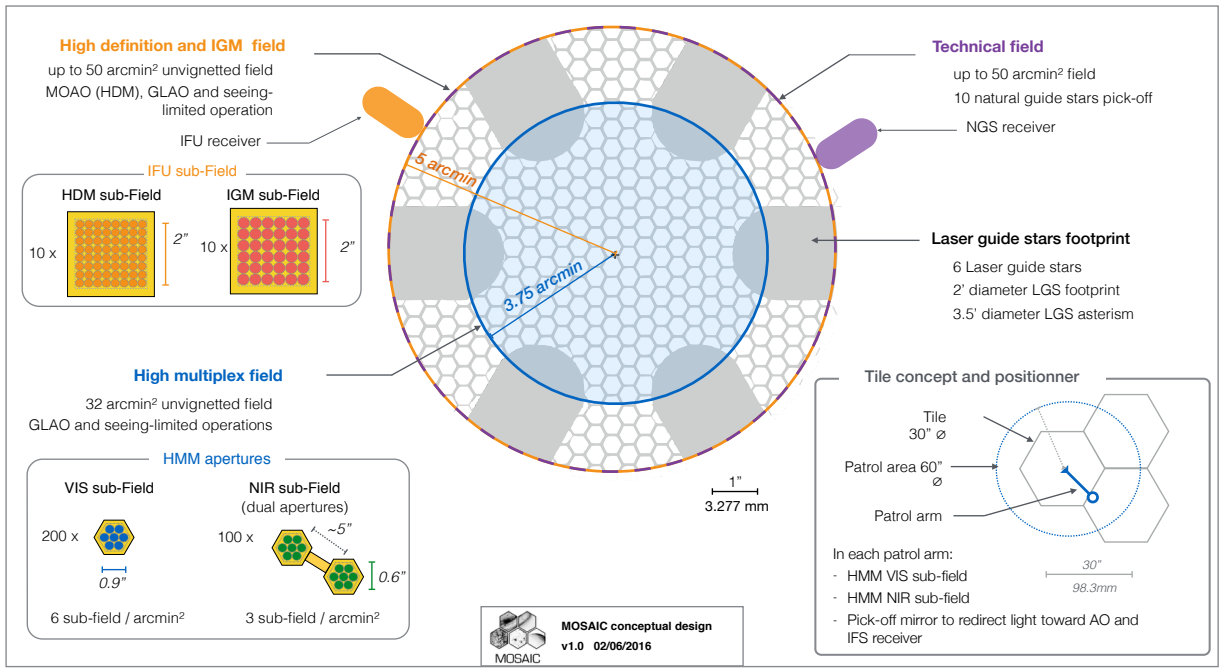

**Figure 3.** Proposed implementation of the focal plane using tiles to correct the non-telecentric design of the telescope (see [7] for more details).

## 3.6 NIR detectors and saturation by OH sky lines

The saturation of strong skylines is a well-known limitation in near-infrared observations. This is especially critical in bands longwards of 1.4μm, where skylines are almost 10 times more intense than in the Y and J band (see Figure 2). The presence of these strong skylines has limited typical Detector Integration Time (DIT) values in the near-IR to 900s on 10m-class telescopes. Since the E-ELT has a collecting area 20 times larger, MOSAIC observations (when using the same spatial sampling) will be affected by the fast saturation of skylines. We are investigating (see [7] for more details) the impact of strong skylines on the maximum MOSAIC exposure time and on MOSAIC operations. For instance, preliminary calculations in the HMM mode indicate that exposures will have to be limited to ~55-74 s in the H band depending on spectral resolution. The relatively short DITs required in the J- and H band to avoid saturation is problematic in terms of signal-to-noise and overhead. The major issue is that such short individual exposures become limited by the read-out noise (RON) in the J- and H band.
 In this regime, the penalty in terms of signal-to-noise ratio scales with the square root of the number of NDIT. This means that the final signal-to-noise of an 8 x 60s observation is degraded by about a factor 3 compared to a single exposure of 500s. Moreover, the additional readout time in the H band increases the overhead. To expose for 500s on target requires splitting observations into 8 DIT~60s exposures in the H band, which translates into 48s additional overhead from the read-out (assuming a read-out overhead of 6s).

These calculations have to be confirmed but they already suggest that alternative read-out modes and spectrograph trade-offs should be investigated to prevent saturation and RON-limited observations (see below).

If confirmed, it leads to severe saturation problems especially around 1.6 μm and this may lead to reduce individual exposures to values as low as 30 sec to avoid image persistence on the detector and then alter the efficiency of all E-ELT NIR spectrographs.

The proposed solution is to calibrate the location of the skylines and use the window-read mode to zero the affected pixels, thus allowing suitably long exposure times (see [8]). This will be discussed with ESO in due course.

## 4. CONCLUSION

At the consortium level, meetings take place on a monthly basis. It allows a wide, representative discussion of all MOSAIC aspects (science, instrument technical progress, communication on the MOS instrument, etc.).

A System Meeting is organized every two weeks to check the progress of the Phase A study and to discuss technical issues at each and every WP level. It finally allows progress on trades that need to be completed before the definition of an overall instrument architecture. Adaptive Optics is a specific WP that requests a huge effort at the very beginning of the Phase A since it will drive a lot of technical choices within other WPs. As such, an internal AO WP kick off meeting was organized in Paris (April 8, 2016) to both decide what simulations are needed and how the work will be shared. It has also been decided that an AO meeting will take place on a monthly basis to check progress and discuss issues.

The MOSAIC Board has approved the Memorandum of Understanding (MoU) for the Phase A, which is entitled "Phase A Study of the E-ELT-MOS instrument and definition of Public Surveys". The Steering Committee that includes all the MOSAIC partners has also met and is strongly supporting MOSAIC, including the investigation to raise a significant cash amount. ESO is facing a probable hardware cost problem for the E-ELT MOS, where the 18 M€ cap in the call for proposals could lead to a severe limitation in its capacity. Hardware cost estimates from earlier studies, performed during the last round of phase A studies, generated a provisional hardware budget near 30 M€ to reach a multiplex of ~ 200. The currently estimated multiplex within the proposed cost cap could be as low as 70. The latter number would severely limit survey speeds and effectively move costs from the instrument hardware into very expensive telescope time (a false economy). A compromise has been agreed upon between ESO and the MOSAIC Consortium. ESO will commit to examine the situation before the end of Phase A, including (i) a costed design showing which capabilities an 18M instrument will have and which it will miss, (ii) an idea of what the additional capabilities will cost, and (iii) a clear picture of which key science cases will be driving the specifications.

In summary, we anticipate that MOSAIC will become an extremely productive and heavily over-subscribed instrument, providing high-impact science results for years to come across a broad range of topics in contemporary astronomy. Given the present status of the E-ELT Instrumentation Plan, MOSAIC [9] will be the only E-ELT instrument able to produce the sample sizes needed for significant progress in cosmology, extragalactic and Galactic domains beyond the Galactic Plane, getting spectra of a significant number of targets among the thousands included in one field of view. In particular, MOSAIC will provide a world-leading and unique capability for spectroscopic follow-up of the most distant galaxies in the Universe, exploiting the legacy of the JWST mission.

**ACKNOWLEDGEMENTS:** We are very grateful to the whole MOSAIC team and consortium for their work towards the future E-ELT Multi Object Spectrograph. Besides the authors, the MOSAIC team also includes Gerard Rousset (Observatoire De Paris), Richard Myers (Durham Univ.), Olivier Le Fèvre (Lam, Marseille), Alexis Finogenov (Helsinki Univ.), Bruno Castilho (LNA, Itajuba), Goran Ostlin (Stockholm Univ.), Jesus Gallego (Madrid, Computense Univ.), Fabrizio Fiore (Inaf-Osservatorio Astronomico Di Roma), Bodo Ziegler (Vienna Univ.), Jose Afonso (Ia, Lisbon Univ.), Marc Dubbledam (Durham Univ.), Phil Parr Burman (STFC-UK-ATC), Tim Morris (Durham Univ.), Tristan Buey (Observatoire de Paris), Fanny Chemla (Observatoire de Paris), Eric Gendron (Observatoire de Paris), Andreas Kelz (AIP, Potsdam), Isabelle Guinouard (Observatoire de Paris), Ian Lewis (Oxford Univ.), Kevin Middleton (STFC-RALSPACE, Oxford), Ramon Navarro (NOVA), Marie Larrieu (IRAP, Toulouse), Thierry Contini (IRAP, Toulouse), Kjetil Dohlen (LAM, Marseille), Niklas Harald (Goettingen Univ.), David Le Mignant (LAM, Marseille) and Yanbin

Yang (Observatoire de Paris). A special and warm thank is also given to the Science Team and to the whole community of scientists supporting the MOS in Europe and in other countries.